\documentclass[a4paper,fleqn,usenatbib]{mnras}
\usepackage{newtxtext,newtxmath}

\usepackage{amsmath}
\usepackage{amsfonts}
\usepackage{amssymb}
\usepackage{calrsfs}
\usepackage{wasysym}
\usepackage{esint}
\usepackage{color}
\usepackage{soul}
\usepackage{graphicx}
\usepackage[T1]{fontenc}
\usepackage{aecompl}

\newcommand\rmp{Rev. Mod. Phys.}

\title[SNR Size Distributions \& the ISM]{Inferring Properties of the ISM from Supernova Remnant Size Distributions}

\author[B. Elwood et al.]{
Benjamin D. Elwood,$^{1}$\thanks{bde14b@my.fsu.edu, jwmurphy@fsu.edu}
Jeremiah W. Murphy,$^{1}$\footnotemark[2]
Mariangelly D\'iaz-Rodr\'iguez$^{1}$
\\
$^{1}$Department of Physics, Florida State University, Tallahassee, FL 32306\\
}

\date{Accepted XXX. Received YYY; in original form ZZZ}

\pubyear{2017}

\begin{document}
\label{firstpage}
\pagerange{\pageref{firstpage}--\pageref{lastpage}}
\maketitle

\begin{abstract}

We model the size distribution of supernova remnants to infer the
surrounding ISM density.  Using simple, yet standard SNR evolution
models, we find that the distribution of ambient densities is remarkably narrow; either the
  standard assumptions about SNR evolution are wrong, or observable
  SNRs are biased to a narrow range of ambient densities. We show that the size
distributions are consistent with log-normal, which  severely limits the number of model parameters in any SNR
  population synthesis model. Simple Monte
Carlo simulations demonstrate that the size distribution is indistinguishable
from log-normal when the SNR sample size is less than 600. This implies
that these SNR distributions provide only information on the mean and
variance, yielding additional information only when the sample size
grows larger than $\sim{600}$ SNRs. To infer the parameters of the ambient density, we use
  Bayesian statistical inference under the assumption that SNR
  evolution is dominated by
    the Sedov phase. In
particular, we use the SNR sizes and explosion
energies to estimate the mean and variance of the ambient
medium surrounding SNR progenitors. We find that the mean ISM particle density
around our sample of SNRs is
$\mu_{\log{n}} = -1.33$, in $\log_{10}$ of particles per
cubic centimeter, with variance $\sigma^2_{\log{n}} = 0.49$. If interpreted at face value, this implies that most SNRs result from supernovae propagating in the warm, ionized medium. However, it is also likely that either SNR evolution is not dominated by the simple Sedov evolution or
SNR samples are biased to the warm, ionized medium (WIM).
\end{abstract}




\section{Introduction}
\subsection{Motivation}
The evolution and properties of Supernova Remnants (SNRs) are useful in studying the physics of
the interstellar medium, as well as their respective progenitor
stars. While the evolution of their expansion has been studied for half a century, more recent efforts have
focused on the statistics of their size distributions and progenitor
masses \citep{gog09, bad10, mur11, asv14, jen12, jen14, wil14}. For example,
  one may characterize the age of SNRs, thereby inferring the age of
  the star that exploded and its progenitor mass.  Because SNRs last
  far longer than SNe themselves, there are far more observable SNRs in
a galaxy than SNe.  Therefore, SNR catalogs are important tracers of
recent SNe, for both Ia and CCSN. \citet{bad10}, for example, used SNRs to constrain the delay time distribution for
SN Ia.  \citet{jen12} and \citet{jen14} combined SNR data from M31 and M33 to obtain a distribution of
  progenitor masses for core-collapse supernovae. Using 115 SNRs, they found a clear minimum mass for CCSNe of
  7.3 $M_{\odot}$.
  Interestingly, they also found far fewer high mass progenitors
  compared to the Salpeter IMF. This suggests that either the most massive stars do not explode as
supernovae or that selection effects bias against finding SNRs in the
youngest regions. Given this result, we are motivated to
  understand the environments surrounding these SNRs and infer any
  potential biases in the SNR catalogs.

In this paper, we investigate whether there is a bias for finding
  SNRs in certain environments.  Our approach will be to use statistical
analyses of the SNR size distributions $N(<D)$--$D$ for M31 and M33
 ($N$ being number and $D$ diameter). In principle,
the size of a supernova remnant depends upon explosion energy,
age, and the density of the interstellar medium. Thus,
the distribution of SNR sizes depends upon the distribution of ISM
densities; any bias affecting SNR catalogs might therefore manifest
itself in the distribution of densities.  

Two approaches can be employed to determine whether a significant
  bias exists. One is to model the observations including bias, and
  the other is to model the observations excluding the bias and test its consistency with other observations. If the model
  without bias is shown to be inconsistent with other observations,
  then one can be confident that there must be a significant bias.
  Modeling the bias is often the preferred option; however, this can be complex and adds many unconstrained
  parameters to the modeled data \citep{sar16}. Presently, modeling SNR
  evolution and visibility requires many uncertain model parameters for which there are only a handful of observables (such as diameter and radio luminosity).  Therefore, the first step is to
show that there is a bias, with the simplest and least complex method being to attempt to model the SNR evolution and infer the ISM
density without the bias. If
the resulting inference does not match expectations of ISM density from
other observations and simulations,
then this effectively rules out the model without a bias and suggests
a strong bias.

\citet{sar16} present the most recent and comprehensive attempt regarding these efforts by modeling the radio emission and observable lifetimes of SNRs, a necessary step in understanding the bias in SNR catalogs. However, modeling the SNR radio emission relies heavily on correctly modeling cosmic ray particle production, SNR evolution, cosmic ray feedback, and synchrotron emission. Due to the many uncalibrated parameters in each of these steps, there is a danger that the inferred conditions will be heavily influenced by the assumed priors for each of these parameters; with so many parameters, it will be difficult to falsify the predictions. Nonetheless, as \citet{sar16} demonstrate, modeling the radio emission and observable lifetimes will be important for properly constraining the SNR evolution. Therefore, it will be important to constrain the many parameters in the radio emission models.

While the important work of calibrating the radio emission continues, there is another approach to assessing the bias that is simpler and leads to clear, falsifiable predictions. In this work, we simply model the SNR evolution with the standard assumption that all SNRs are observed in the Sedov phase of expansion with an observable lifetime that is determined by the onset of the radiative phase. The simplicity of the model, due to very few parameters, offers clear and directly falsifiable results. Even in the event that the inferred densities are falsified, we are certain to learn that obvious and standard assumptions about SNR evolution are seriously mistaken. Such a result would greatly illuminate the role of bias in SNR catalogs.

\subsection{Background}
For decades, there have been
several attempts to infer the properties of the ISM based upon SNR
size distributions \citep{fus84, brk86, brk87, bad10, asv14}. This was
motivated, in part, by an early analysis of size distributions in the
Magellanic Clouds by \citet{mat83} in which the authors observed a
linear $N$--$D$ cumulative
distribution, more consistent with the free expansion phase than Sedov phase. Assuming a single explosion energy and ISM density, this
is at odds with a distribution ($N \sim D^{2.5}$) expected
  for the Sedov phase \citep{mat83}.  Instead, a linear cumulative
  distribution is most consistent with the free expansion phase of evolution. Considering that
free expansion takes place within a short time frame ($t_{FE} \sim 10^2$ yr), this interpretation seems unlikely.
  For example, the short timescale for the free expansion
  phase would suggest far fewer SNRs per galaxy than are actually
  observed.  The SN rate for a typical $L^{*}$ galaxy is 1/100 years
  \citep{cap99,gig06,li11}, implying that if free expansion phase
    dominates, one should only observe $\sim 1 - 10$ SNRs per galaxy at any given
time.  Yet, SNR catalogs for nearby galaxies contain about 100 SNRs, which is more consistent if one considers the longer Sedov phase.

Shortly after the findings presented by \citet{mat83}, \citet{huh84} argued that those proposing a free expansion interpretation did not carefully consider how selection effects, explosion energies, or ISM densities could influence the diameter distributions and instead showed that a linear relationship was more consistent with Sedov evolution. No doubt influenced by this work, more contemporary analyses were provided for both M33 and M82 by \citet{ban10} and for the Small and Large Magellanic
Clouds by \citet{bad10}. \citet{ban10} state that a linear cumulative distribution does not imply free expansion, if caused by a combination of SNR data expanding into different ambient densities. Similarly, since SNRs in free expansion can
experience shock velocities up to $\sim 10^4$ km/s, those expanding to
a diameter of up to 60 pc (a maximum first obtained by \citet{mat83}) would have thousand-year timescales. This result leads to
both a highly-overestimated SN rate (as discussed earlier) and low-limit progenitor mass when
compared to SNR data from any known galaxy \citep{mao10}. In
light of these contradictions, contemporary views hold that most
observed SNRs are in the Sedov phase of expansion \citep{lee14}.

\citet{fus84} showed that the SNR
data could be replicated with Sedov theory by rejecting a large-scale
homogeneous ISM and incorporating a spread of densities in their calculations. Additionally, the completeness of SNR data became an increasingly important factor in whether or not erroneous
assumptions were made regarding free expansion. \citet{brk86, brk87} continued this investigation by recognizing
a difficulty in ``proving" SNRs to be in free expansion: SNR diameter distributions are statistically vulnerable to outside influences from ambient density or explosion
energy. Thus, a supporting expansion law can be obscured by any variation in parameters that randomize the SNR diameters. Additionally, via Kolmogorov-Smirnov analyses, the
author concluded that neither a free expansion nor Sedov expansion
power law could be satisfactorily rejected, based upon a lack of completeness in the SNR data \citep{brk87}. As SNR data has become more readily available, due to
surveys such as \citet{Lee14} for M31 and \citet{lon10}
for M33, this problem has been somewhat ameliorated and support for a free
expansion observation has declined. As we will
demonstrate shortly, diameter distributions appear to be closer to log-normal in nature, rendering a linear distribution only approximately valid within a certain range.

A more recent motivation for studying SNR distributions was provided by \citet{bad10}, whose analysis included a discussion of the SNR size distributions for the SMC and LMC. Having gathered 77 multi-wavelength SNRs, the authors argued for an approximately-linear characterization up to a diameter of 60 pc, stating that this behavior is determined by the physics of the SNR interaction with the interstellar medium. Their model suggests that SNRs expanding beyond this diameter exit the Sedov phase of expansion to the radiative, quickly fading from detection with a lifetime that depends upon the surrounding ISM density. The authors note that a linear size distribution can arise naturally if the distribution of ISM densities follows a power law with an exponent of -1 with a wide range of densities.

Analyses of this nature suggest an important hypothesis: size
distributions of supernova remnants appear to hold clues as to the ISM
environments in which they propagate. Motivated by this possibility,
we have initiated a statistical analysis of our own, thanks to
external data provided by \citet{Lee14} for M31 and \citet{lon10} for
M33. As we show in section \ref{datastats}, we note that these size distributions are log-normal,
which is a natural consequence of Sedov expansion. From this, we derive a distribution of ISM
densities, allowing us to extract useful statistics on the SNR
environments by applying the Central Limit Theorem.

In order to understand the size distribution of SNRs, it is essential
to understand the dynamics of SNR evolution, particularly
lifetimes. It is often assumed (but not explicitly verified) that the
dominant, observable phase of an SNR is given by the Sedov-Taylor
solution. This is because the observable timescales of both the free
expansion ($t_{FE} \sim 10^2$ yr) and radiative phase are too short to
be adequately detected. Considering that the cooling time for a given ISM density goes as $t_{cool} \sim {\rho}^{-1}$ \citep{tru98, blo98}, SNRs in the radiative phase can quickly fade
from view, especially if merging with dense environments. Entry into the radiative
phase allows for brief X-ray detection, after which the threshold
becomes too low for reliable observation, despite the fact that
radiative lifetimes can similarly last for thousands of years \citep{cio88, blo98}. Therefore, we assume the maximum
time to observe an SNR to be approximately equal to the maximum
{\emph{Sedov}} lifetime, at which the cooling lifetime becomes equal
to the expansion time for a SNR:
\begin{equation} \label{eq:tmax} t_{max} \sim R_{max}^{5/2}{\rho}^{1/2}E^{-1/2} \end{equation}
 We derive this in the equation for its namesake radius, up to a dimensionless constant:
\begin{equation} \label{eq:Sedov} R \sim E^{1/5}t^{2/5}{\rho}^{-1/5} \end{equation}
where $E$ is the explosion energy of the progenitor star, often being on the order of $10^{51}$ ergs.

Assuming the Sedov phase to be the simplest and most natural phase affords us a useful avenue to proceed. We analyze the statistics of SNR size distributions to infer properties of the SNR Sedov variables, such as
the ambient density. As will be shown in the next section, the size
distributions for radii closely match a
log-normal distribution for our sample sizes ($N
\sim 10^2$). Interest in this observation stems, in part, from the
Central Limit Theorem (CLT) which states, roughly, that the average of
a sum of a large number of random, independently-distributed variables
will be approximately normal, regardless of the underlying
distribution. By extension, when the logarithm of a product of such
variables approaches a normal distribution, the product itself is a
log-normal distribution. For the Sedov radius, eq.~\ref{eq:Sedov}:
\begin{equation} \label{eq:logSedov} \log{R} \sim \frac{1}{5}\bigg(\log{E} + 2\log{t} - \log{\rho}\bigg) \end{equation}
log-normality exists conditionally if the energy, ISM density, and SNR
lifetime are independently distributed and have well-defined means and
variance. In fact, it is this condition that bears the most relevance
to this paper; conversely speaking, if a log-normal distribution is
obtained from some sample, one is able to extract statistically
meaningful information on the mean and variance of the random
variables. For the purpose of our study, we seek to extract such
information on the ISM density to see if SNR surveys are biased
towards particular environments.  
  
Considering that the Central Limit Theorem holds only for many random, independent variables, it is clear that our analysis will be valid only within a certain criteria. The Sedov radius contains only three variables (energy, density, lifetime), with the lifetime being {\emph{dependent}} upon explosion energy and density. This likely plays a role in our simulated distributions being inconsistent with log-normal above a certain sample size, roughly 600 SNRs (to be shown later). However, statistics suggest that our simulations are consistent with the M31 and M33 data below this threshold, thereby making our analysis approximately valid within the CLT context.

This paper begins by presenting size distributions of previous M31 \citep{lee14} and
M33 data \citep{lon10}, for a total of 187 SNRs. We suggest in Section 2 that a log-normal
relationship for the diameter (or radius) is most appropriate, in
contrast to previous attempts characterizing it as linear, such that
we are naturally led to considering the statistical consequences of
the Central Limit Theorem for the Sedov variables. We then proceed, in Section 3, to
Monte Carlo simulations of the Sedov radius by modeling the blast
energy, SNR lifetime, and ambient density as random variables. We see that, for $N \sim 10^2$, a log-normal distribution is similarly
obtained, to be lost only when $N$ grows large. To acquire more
detailed information, we then employ a Bayesian inference scheme \citep{for13}, assuming
a Gaussian log-likelihood, for the purpose of obtaining means and
variance for the Sedov variables. Our central result is that our
analyses point to an intriguingly low variance for the ISM density
($\sigma^2 = 0.49$); this suggests that SNR surveys are presently biased
towards rather uniform environments, most likely the warm, ionized medium. These results are consistent with other recent efforts suggesting narrow density distributions \citep{gat16, sar16}. The reason for such is unknown,
though we offer possible explanations in our discussion, one being
that the lower-end of size distributions suffers a cutoff resulting
from stellar cavities formed from pre-explosion solar wind of the
progenitor star. We propose alternative corrections to our Sedov model to account for SNR interactions with the stellar wind and surrounding ISM. Our findings are then summarized in the conclusion.
\section{Data and Statistics} \label{datastats}

\subsection{SNR Data}

One of our primary motivations in considering the size distribution of
SNRs is to assess whether there are biases in the SNR catalogs used by \citet{jen12, jen14}.  Therefore,
in this analysis we focus on those catalogs, which includes SNR data
from M31 and M33. The M31 data
includes a fairly new, optically selected survey by \citet{Lee14}, inspired by previous efforts from \citet{brn93, mag95, wil14} and using data from the Local Group Survey \citep{mas06}. For both
M31 and M33, SNR identification was done using [SII]:H$\alpha$
ratios \citep{gor98}; objects identified with ratios exceeding 0.4 fall into the
category of SNRs. Then, to verify the SNRs, these authors used radio
and X-ray data to confirm their SNR status \citep{Lee14, lon10}.

\citet{lon10} considered it likely that observational and physical selection effects contribute to the shape of SNR distributions. From an observational standpoint, the authors mention a tendency to focus on isolated objects, whose shells are easily distinguishable, as well as a bias towards objects whose optical diameters were below 10" (a result of data sensitivity to beam diameter). As detection probabilities for large-diameter SNRs are lower than small-diameter ones, this is always an effect to consider when evaluating SNR data sets. However, \citet{jen14} suggests that selection effects related to detection methods do not significantly contribute to SNR catalogs, though they admit to leaving this as a somewhat open-ended question. In addition to observational biases, \citet{lon10} also discusses the aforementioned physical effect in which SNRs that explode into dense ISM environments will dissipate much more rapidly, escaping detection (see eq.~\ref{eq:cooling}). \citet{jen14} likewise mentions the possibility that any SNR cataloger biased towards particular progenitors or ISM environments could affect the overall distribution shape. Thus, understanding the nature of the bias involved in the relationship between SNRs and the ISM is not only of theoretical interest, but is a realistic problem in SNR observation.
  
  It is important to mention that this work provides no distinction between CCSN and Type Ia SNR
  candidates. Recent works suggest that while CCSNe dominate SNR
  samples, the fraction of Type Ia SNe are not negligible and
  typically constitute roughly a quarter of total SNe \citep{li11,
  Lee14, sar16}. \citet{jen14} in particular provide support for
  ~25\% being an upper limit on Type Ia identification, both due to
  differences in surface brightness, as compared to CCSNe, and spatial
  distributions. Eventually, one should make a distinction
  between SNRs from CCSNe and SN Ia, as there could be a difference in the
  ISM properties associated with each. \citet{sar16} attempted to model the SNR size distribution including both
  SNIa and CCSNe, but this requires yet another uncertain parameter, the
  fraction of SNIa compared to total.  This fraction is very difficult
to determine from the properties of SNRs in external galaxies.  The
best method is to age-date the surrounding stellar population for each
SNR, an analysis that will come in due time. Until then, we are forced
to treat the entire SNR population as a whole.  Therefore, we will be
inferring the surrounding ISM density distribution for both CCSNe and
SNIa.

\subsection{Size Distributions}

\begin{figure}
\centering
{\includegraphics[width=0.45\textwidth]{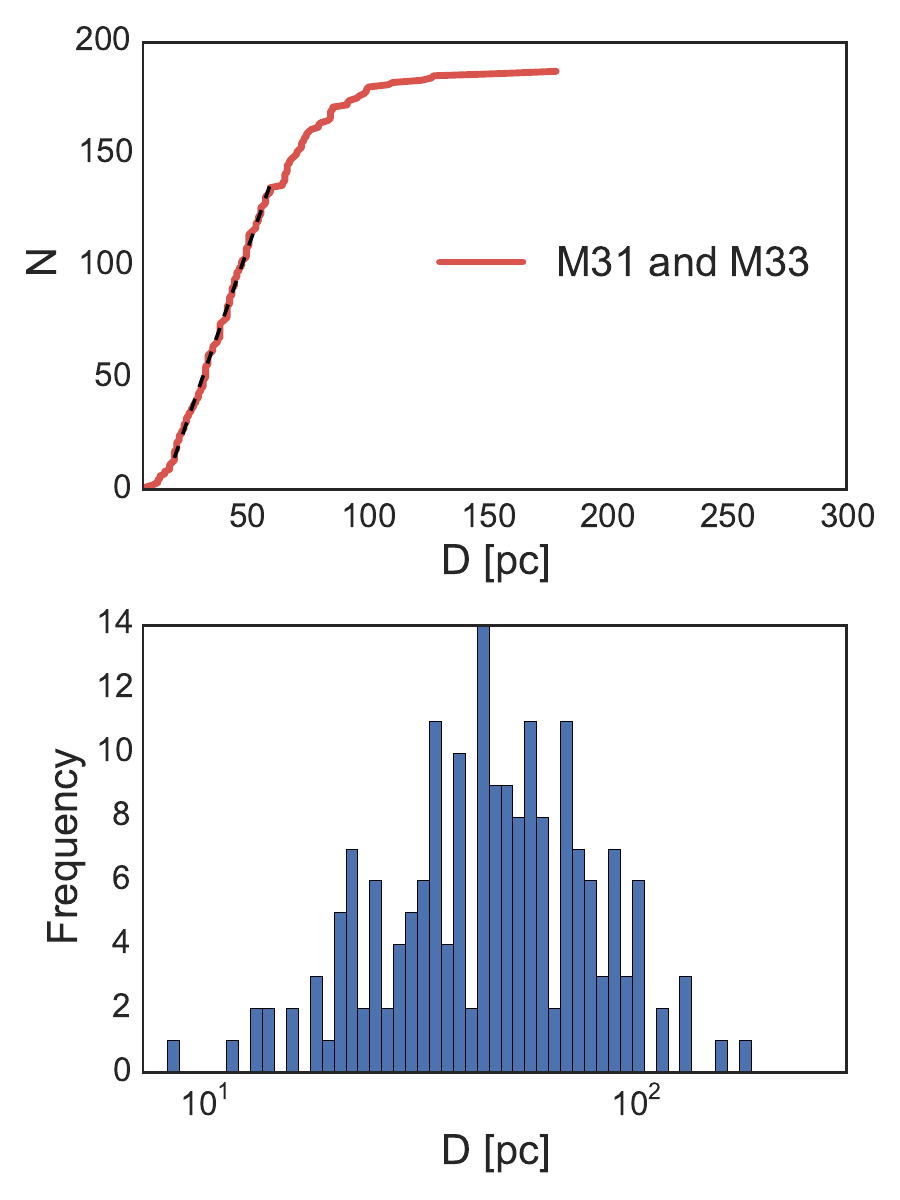}
  \caption{(above) Cumulative distribution for both M31 and M33 SNR
    data. Dashed line indicates linear characterization for SNRs up to
    the ``cutoff diameter" of 60 pc. Ignoring the rest of the data leads to assuming a linear distribution and the erroneous interpretation of free expansion. (below) Histogram for M31 and M33 in log space; log-normality is present with P-value $\approx 0.277$, a consequence of Sedov phase expansion.}
      \label{fig:m31dist}
  }
\end{figure}

Figure ~\ref{fig:m31dist} presents the size
  distributions for M31 and M33 SNRs; the top panel shows the
  cumulative distribution and the bottom panel shows the frequency distribution. By plotting the frequency distribution in the
log, it is easy to see that the size distribution is consistent with a
log normal distribution. To allay statistical concerns with mixing our M31 and M33 samples, we performed a simple Kolmogorov-Smirnov test between the two sets to obtain a P-value of $\approx 0.146$ to verify that both samples could come from the same underlying distribution, thereby justifying a combined data set.

In the past, others assumed that linear distributions implied a detection
of SNRs in their free expansion phase, though, as mentioned
previously, this view is no longer shared. A simple reason for this is
that the timescale for free expansion can only last for a hundred or
so years. The transition from free expansion to the Sedov
  phase begins when the swept-up mass roughly equals the ejecta mass,
  $M_{\rm ej} \sim M_s \sim R^3 \rho$; with a blast energy of $10^{51}$ erg, progenitor mass of $8
M_\odot$, and an ISM particle density of 0.1 cm$^-3$:
\begin{equation} \label{eq:FEtime} t \sim E^{-1/2}M^{5/6}{\rho}^{-1/3}
  \sim 10^2 [yr] \end{equation}
 Assuming free
expansion also contradicts the observed SNR data for the Magellanic
Clouds, now believed to be in the adiabatic (Sedov) phase of expansion
\citep{bad10}. In this work, we not only assume SNRs to be in the
Sedov phase of expansion, but propose that a Gaussian distribution
fully encapsulates the observed data, motivated largely by the
obviously log-normal shape of the histogram in Figure~\ref{fig:MC300}.
\subsection{Density Distribution Statistics}

We can easily model the distributions of dependent variables by approximating the SNR distributions as log-normal. By doing so, we extract meaningful statistics of these variables, such as
inferring properties of the ISM. First, recall the Central Limit
Theorem, which states that, for a sufficiently large number of random,
independent variables (with well-defined means and variance), the mean
of a large number of iterates will be normally distributed. By extension, when the logarithm of a product of such variables (taking positive values only) approaches a normal distribution, the product itself is log-normally distributed.

We extend these statistical properties to the Sedov radius (eq.~\ref{eq:Sedov}), a product of energy, time, and interstellar
density. Since the SNR diameters are log-normally distributed, we can
extract statistical information on the mean and variance, specifically the ISM density. We first note that the variance of our given data for $\log{R}$ is quite small; it then becomes easy to calculate a quick estimate of the maximum variance for $\log{\rho}$ if we assume the Sedov variables to be independently-distributed. For the Sedov radius, our mean and variance (in the log) are:

\begin{equation} \overline{\log{R}} = \frac{1}{5}\bigg(\overline{\log{E}} + 2\overline{\log{t}} - \overline{\log{\rho}}\bigg) \end{equation}

\begin{equation} \sigma^2_{\log{R}} = \bigg(\frac{1}{5}\bigg)^2\sigma^2_{\log{E}} + \bigg(\frac{2}{5}\bigg)^2\sigma^2_{\log{t}} + \bigg(\frac{1}{5}\bigg)^2\sigma^2_{\log{\rho}} \end{equation}

To calculate an upper bound on the density in the variance, we set the other two variables to zero. Then, we obtain the largest possible variance for the log-density:

\begin{equation} \sigma^2_{\log{\rho}} = 25\sigma^2_{\log{R}} \end{equation}

Observationally, the variance for the radius
is $\sigma^2_{\log{R}} \approx 0.055$, thus
setting an {\emph{upper limit}} for the density variance
$\sigma^2_{\log{\rho}} = 1.375$. This exercise is indeed simplified,
for cross-terms such as the lifetime depend upon, for example, the
blast energy and ISM density. Nonetheless, the exceptionally small
variance suggests that a more thorough analysis,
  which we present in the next section,
  will produce an exceptionally narrow distribution for densities.

\section{Simulations}
We seek to infer the distribution of densities surrounding the SNRs by invoking a model that translates the distribution of densities into a distribution of SNR sizes. We use Monte Carlo simulations modeling the explosion energy $E$, progenitor mass $M$, and ambient ISM density $\rho$ as random variables in the form of power laws:
\begin{equation} \rho = {\rho}_+^R{\rho}_-^{1-R} \end{equation}
\begin{equation} E = E_+^QE_-^{1-Q} \end{equation}
\begin{equation} t = S\big(t_+ - t_-\big) + t_- \end{equation}
\begin{equation} \bigg(W\Big(M_+^{-1.35}-M_-^{-1.35}\Big)+M_-^{-1.35}\bigg)^{-1/1.35} \end{equation}
where $Q$, $R$, $S$, and $W$ are random numbers that range from 0 to 1
and the $+$ and $-$ indicate maximum and minimum values, respectively (with progenitor mass ranging from one to one hundred solar masses). We now consider the SNR cooling time for a given ISM density:
\begin{equation} \label{eq:cooling} t_{cool} \sim \frac{kT}{\rho{\Lambda}(T)} \end{equation}
with $\Lambda{(T)}$ being the cooling function as provided by \citet{tru98} and \citet{blo98}. Assuming a Sedov model in conjunction with eq.~\ref{eq:cooling} leads to the maximum radius as found in \citet{ban10}:
\begin{equation} \label{eq:tepsilon} R_{max} \sim E^{(3-2\epsilon)/(11-6\epsilon)}{\rho}^{-(5-2\epsilon)/(11-6\epsilon)} \end{equation}
after which the SNR enters the radiative phase and fades from view. The exponent $\epsilon$ arises from assuming a cooling function of the form $\Lambda(T) \sim T^{\epsilon}$ as found in \citet{tru98} where, for our simulations, $\epsilon = 0.5$.

\begin{figure}
\centering
{\includegraphics[width=\columnwidth]{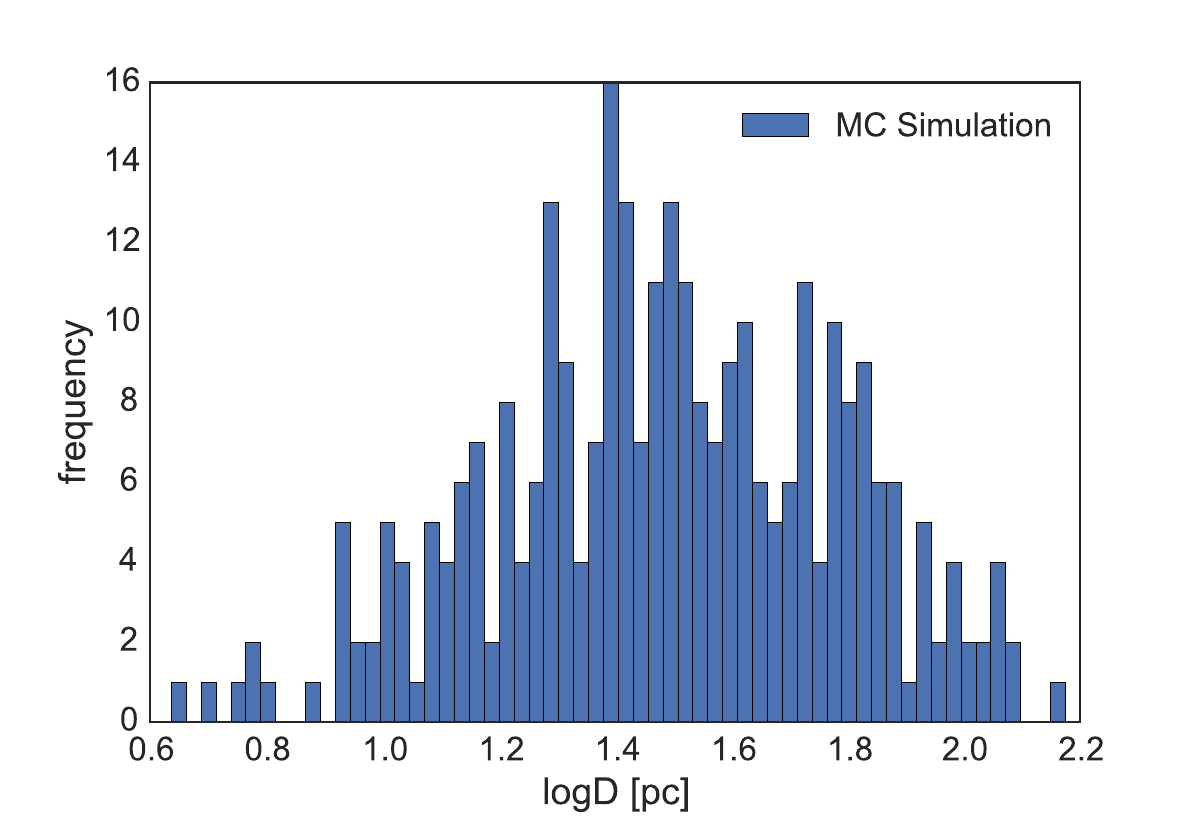}}
\caption{Monte Carlo scheme for Sedov radius (in the log), for
  N=300. Log-normality is therefore well-produced in the theoretical
  framework of the simple Sedov radius. P-value $\approx 0.23$.}
    \label{fig:MC300}
{\includegraphics[width=\columnwidth]{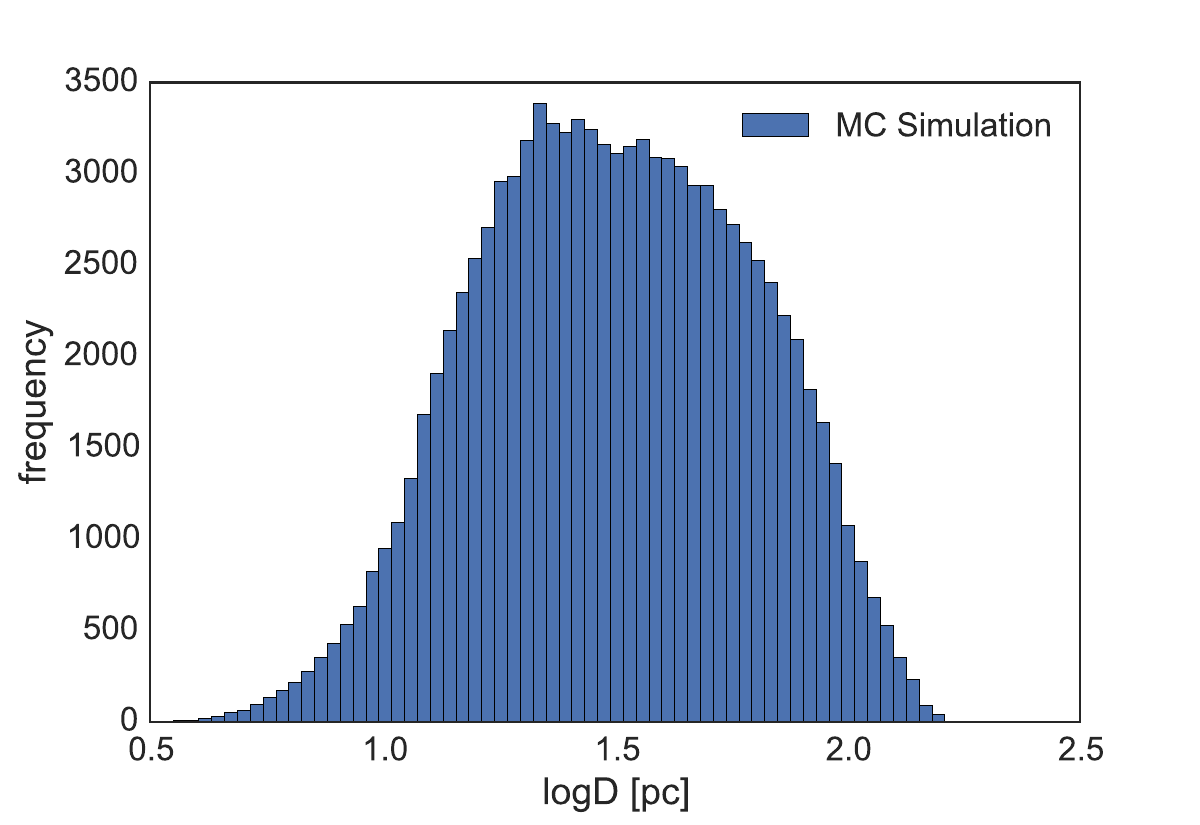}
\label{fig:MC100000}}
\caption{Monte Carlo scheme for Sedov radius (in the log), for
  N=100000. Note the loss of log-normality when sample grows
  arbitrarily large, thereby granting additional information on the statistics, such as kurtosis or skewness (clearly present).}
{\includegraphics[width=\columnwidth]{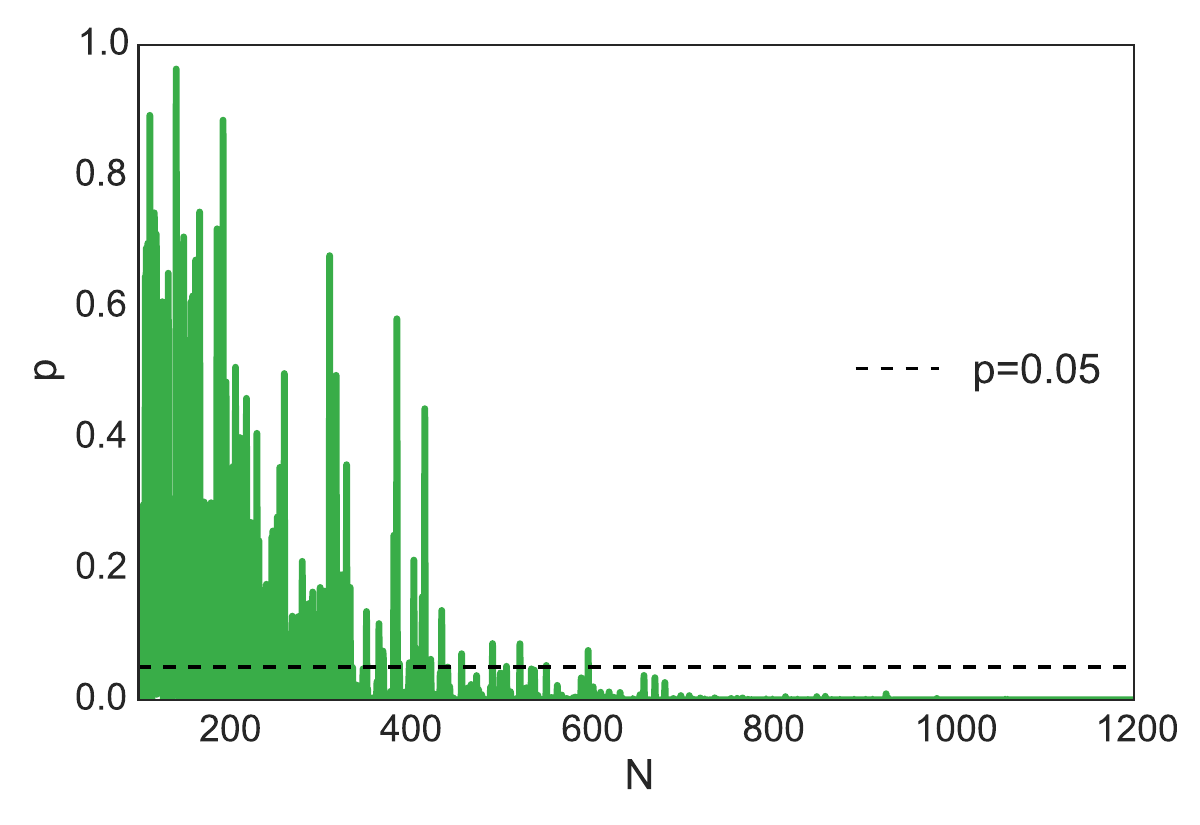}}
\caption{P-values versus $N$. Distribution becomes inconsistent with log-normal for
  sample sizes larger than 600.}
  \label{fig:Ptest}
\end{figure}

The result of these Monte Carlo simulations is displayed in Figure~\ref{fig:MC300}, where we are able to reproduce the
log-normal behavior for $N=300$, with $N$ being the number of
simulated SNRs. However Figure~\ref{fig:MC100000} shows that the size distributions are clearly non-Gaussian for large $N$.  The
    central limit theorem states that the distribution tends toward a
    Gaussian when there are a large number of variables. In our simplified model, there are
    only three in the equation for the Sedov radius, so it is somewhat
  surprising that the distribution of sizes is indistinguishable from
  log-normal with a modest number of SNRs ($\sim$600).  For catalogs
  with a large number of SNRs, one will be able to distinguish between
  a normal distribution and a non-normal distribution.  If the
  distribution is \emph{not} normal, then one will be able to constrain more
  parameters and physics about the underlying distributions, rather than merely obtaining information on the mean and variance of the underlying variables.  To
  estimate how large the SNR catalogs under a simplified Sedov model will have to be in order to
  distinguish from a log-normal distribution, we also include a plot
  of p-values for each simulation in which $N$ grows steadily larger
  in Figure~\ref{fig:Ptest}; as indicated, normality ($p \geq 0.05$) is lost at a sample size over 600.

A possible reason for the loss of consistency with log-normal stems from recalling that the
CLT holds only for as long as the variables remain independently
distributed. Clearly, our few variables for the Sedov radius do not
meet the criteria as ``sufficiently large" and, as shown in eq.~\ref{eq:tmax} and \ref{eq:FEtime}, the minimum and maximum lifetimes actually {\emph{depend}} upon mass and blast energy. Thus, while log-normality is clearly present for samples containing a few hundred SNRs, the fact that the Sedov radius ultimately fails the requirements for the CLT to hold suggests that there should be no reason for this to universally hold for all sample sizes.

\begin{figure}
\centering
{%
  \includegraphics[width=0.4\textwidth]{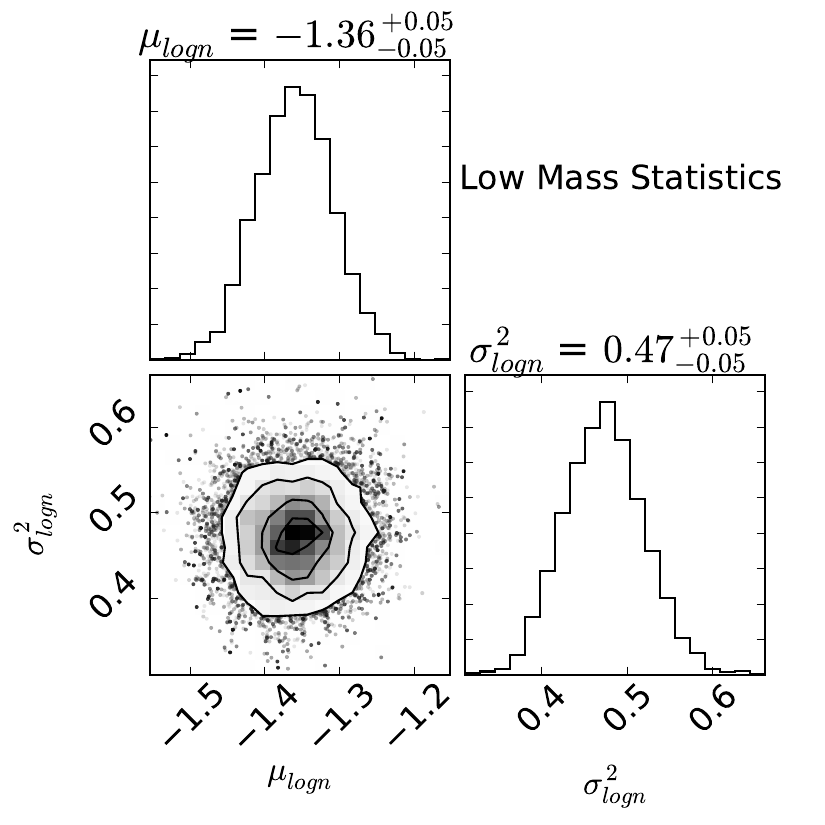}%
  }\par\medskip
\caption{Mean particle density $\mu_{\log{n}} = -1.36$ and variance
  $\sigma^2_{\log{n}} = 0.47$. The small variance in these simulated distributions appears to be consistent with the analysis of the SNR data. \label{fig:lowmass}}{%
  \includegraphics[width=0.4\textwidth]{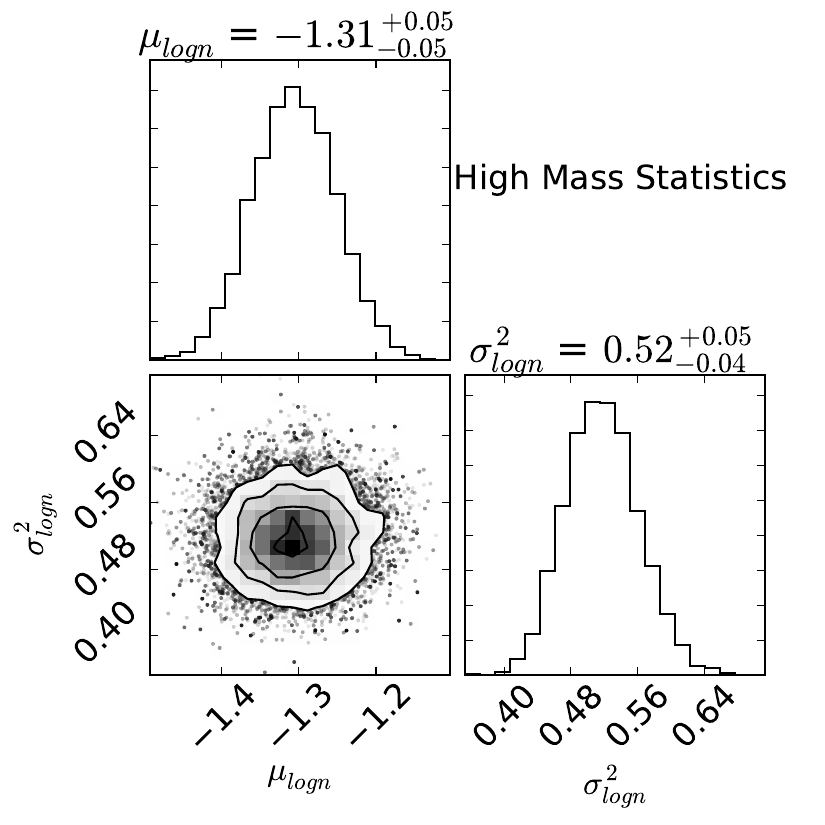}%
  }\par\medskip        
\caption{Mean particle density $\mu_{\log{n}} = -1.31$ and variance $\sigma^2_{\log{n}} = 0.52$. Variance is evidently insensitive to SNR progenitor mass.}
\label{fig:highmass}
\end{figure}

Current SNR catalogs are well below $\approx 600$, so it is reasonable
  to assume that the distribution is log-normal.  Furthermore, since
  the size distribution is consistent with being log-normal we can
  only know about the mean and variance for the distributions of $E$, $t$, and
$\rho$. Therefore, via Bayesian statistical inference, we aim to
infer the mean and variance in the log of the ISM density. Recall that Bayes' theorem states:
\begin{equation} P(M \vert D) = \frac{P(D \vert M)P(M)}{P(D)} \end{equation}
where $P(M \vert D)$ is our {\emph{posterior}}, $P(D \vert M)$ our {\emph{likelihood}}, $P(M)$ our {\emph{prior}}, and a normalization constant $P(D)$. For our purposes, we assume a flat prior such that $P(M \vert D) \propto P(D \vert M)$. In this context, $M$ refers to the model parameters and $D$ the data, such that we seek to calculate the probability of $M$, given $D$. Considering the log-normality of the Sedov radius, our likelihood function is thus assumed to be the log of a normal distribution function with model parameter $\log{R}$:
\begin{equation} \log{P} = -\frac{1}{2}\log\bigg(2{\pi}\sigma^2_{\log{R}}\bigg) - \bigg(\frac{\big({\log{R}} - \mu_{\log{R}}\big)^2}{2\sigma^2_{\log{R}}}\bigg) \end{equation}
whose parameters depend on the diameter data obtained from the M31 and
M33 samples. From this likelihood, we can perform an analysis of the
ISM density by modeling $\rho$ as random, normal distributions. We constrain the statistics of our modeled distributions by calculating the mean and variance of actual explosion data from observed and physical properties of CCSN as found in \citet{ham03}. SNR lifetimes are calculated via eq. ~\ref{eq:Sedov}, ~\ref{eq:tepsilon},
and ~\ref{eq:tmax} and we perform our Bayesian scheme on the density for both low ($7-9M_{\odot}$) and high
($10-120M_{\odot}$) mass ranges. The results are shown in figures ~\ref{fig:lowmass}
and ~\ref{fig:highmass}, in which we obtain very narrow variances for $\log{\rho}$ for both mass ranges; the mean and variance for combined low and high masses come to $\mu = -1.33$ and $\sigma^2 = 0.49$. Our mean values for the density are consistent with those found in previous models of the WIM, such as \citet{hil08} who concluded that WIM gas follows a log-normal density distribution with $\mu_{\log{\rho}} \approx -1.52$. \citet{brk08} also provide supporting evidence that densities in the diffuse ISM are log-normally-distributed, using dispersion and emission measure data derived from pulsar samples. The authors conclude that this behavior is typical for random, non-linear processes such as turbulence.

Our modeling suggests that the range of densities is narrow; while
this may seem peculiar, there may be underlying, theoretical
explanations that we consider in the next section. Additionally, our
narrow density distribution is empirically consistent with recent efforts by
\citet{gat16}, who studied interactions between supernovae and the multi-phase ISM via FLASH hydrodynamic simulations. As part of this study, the authors calculated density statistics for two runs related
to SN rates in different density environments. They reported a mean
density $\bar{n}_{SN} \approx 0.07$ cm$^{-3}$ with standard deviation
$\sigma_{SN} = 0.9$ for one run, and $\bar{n}_{SN} \approx 0.09$
cm$^{-3}$ with $\sigma_{SN} = 1.1$ for the other; their simulated
results suggest that SNRs generally encounter low-density surrounding environments. This is also supported by \citet{kim16} who report that high-resolution models (via TIGRESS) produce large fractions of cluster SNe in low-density regions ($n < 0.1$ cm$^{-3}$).

Comparisons with \citet{sar16} yield mixed results. Though they do not report the inferred ISM density directly, they do report parameters of their model that go into modeling the density. In their attempt to better understand delay time distributions associated with SNRs, \citet{sar16} sought to model visibility times, thereby requiring an inference of environmental densities. However, rather than simply including the local density as a parameter, they model the density with a smooth disk-like distribution exponential:
\begin{equation} \label{eq:SarDensity} n_0(z) = \frac{N_H}{\sqrt{\pi{z_0}^2}}\exp\Bigg({-\frac{z^2}{z_0^2}}\Bigg) \end{equation}
where $n_0$ is the volumetric density, $N_H$ the HI column density,
$z$ the vertical height at which an SNR explodes in a given spatial
location (in M33 for their purposes), and $z_0$ the scale height for a
given galactic disk (again M33). Therefore, two parameters exist, which they effectively reduce to one by
  including the HI column density observations in their analysis.

As the authors did not report a distribution of densities, we must estimate the mean and variance in $n_0$, given their mean and variance in
$z_0$. However, we first note that their ISM density model might be too
  restrictive.  Equation~(\ref{eq:SarDensity}) is a very smooth density
  distribution and does not account for the multiple phases of the
  ISM. On top of the smooth exponential profile, the real ISM is
  percolated with large regions of under dense hot gas and small-volume, but high-mass regions of cold, dense gas.  There is no way for
their model to capture this inherent variation in the densities of
the ISM. Furthermore, using the HI column density is problematic for
two reasons. One, the HI column density represents one specific phase
and temperature of the ISM, and two, the HI column density is an integral
quantity.  For these two reasons, the distribution in HI column
densities is quite narrow and does not represent the inherent
variation in ISM densities \citep{diam89}.  

We then use their distributions for $N_{H}$ and $z_0$ to
  estimate $\sigma^2_{\log n}$ or $\sigma^2_{n_0}/n_0^2$. Assume that $z$ does not vary significantly, with its mean being nearly zero (an assumption based upon the fact that $z$ is drawn from an exponential distribution). Invoking the definition of variance as a sum of derivatives normalized by $n_0^2$ yields:
\begin{equation} \frac{\sigma_{n_0}^2}{n_0^2} = \frac{\sigma_{N_H}^2}{N_H^2} + 4\frac{z^2\sigma_z^2}{z_0^4} + \bigg(\frac{4z^4}{z_0^4} + \frac{4z^2}{z_0^2} + 1\bigg)\frac{\sigma_{z_0}^2}{z_0^2} \end{equation}
Estimates of these variables can be obtained in a couple ways. We
first estimate the values of $\sigma_{N_H}$, $\mu_{N_H}$, and
$\sigma_{z_0}$ from a parameter space figure within \citet{sar16} in
which $\sigma_{N_H} \approx 1.0 \times 10^{21}$ cm$^{-2}$, $\mu_{N_H}
= 2.5 \times 10^{21}$ cm$^{-2}$, and $\sigma_{z_0} \approx 100$ pc. We
use $z_0 = 200$ pc, a value stated by the authors in a figure of
visibility times versus column density. If we further assume
$\mu_{N_H} \approx N_H$, $\mu_z \approx z$ (in which $z = 0$), and
$\sigma_z^2 \approx z^2$ (by virtue of it being drawn from an
exponential distribution), one obtains a normalized variance of
$\sigma_{n_0}^2/{n_0^2} \approx 0.41$, translating to $\approx 0.077$
for the variance in $\log{n_0}$. This fractional variance is similar to the fractional variance for $N_H$, suggesting that the shape of the volume density distribution is determined by the prior of the authors' column density distribution. 

Though \citet{sar16} did not report values for the variance of $\log{n_0}$, private communication revealed an inferred variance of 0.09, similar to the inference reported in this manuscript (S. K. Sarbadhicary and C. Badenes, personal communication, 2017). The mean densities on the other hand are somewhat different. Figures ~\ref{fig:lowmass} \& ~\ref{fig:highmass} show that our mean inferred density distribution is
  $\mu_{\log_{10}(n)} \approx -1.3 $.  For \cite{sar16},
  $\mu_{\log_{10}(n)} \approx 0.5$, nearly two magnitudes larger.  The
  standard deviation for our distribution is $\sigma_{\log_{10}(n)}
  \approx 0.7$, such that the means are two sigmas apart in $\log_{10}(n)$,
  The standard deviation in our mean is 0.05, so the differences in the means are quite significant.

However, the fundamental differences between our analysis and
  that of \citet{sar16} imply that it is not entirely
  appropriate to strictly compare results. In particular, we
infer the densities from the size distributions and use a very crude
model for the lifetime of observability. On the other hand,
\citet{sar16} use a radio light curve model to infer the
  densities and ignore the sizes. Furthermore, our analysis makes no assumptions about a specific density
  model. Rather, we simply infer density statistics using the variance
  obtained from SNR sizes. Alternatively,
  the density model assumed by \citet{sar16} is that of a
  homogenous disk; as the authors note, disk galaxies in general are
  complicated and inhomogeneous with respect to the ISM, and their
  model is likely too restrictive. It would be useful, in future work, to observe how density statistics change with the assumption of more realistic and complicated models and how such statistics compare with those obtained by the more elementary avenue in our work.

\section{Discussion}

These interesting values suggest a narrow derived density distribution that might lead one to consider the possibility that either surveys of supernova remnants are biased towards the warm, ionized medium or that the evolution of SNRs is not dominated by the simple Sedov radius model. However, an example of a non-Sedov model may include the stellar cavities formed before the progenitor explosion via stellar wind. For
some time, it has been known that wind-driven mass-loss plays a
significant role in the evolution of stars, particularly for massive
stars whose end fates result in core-collapse supernovae \citep{chi86, mae00, woo02, smi14}. Strong radiation pressure launches powerful winds that push the interstellar medium out several parsecs
away from the star, forming a pre-explosion cavity between the
progenitor and a shell of ISM material. Upon explosion, a minimum ejecta mass of $6.6 M_{\odot}$ easily overwhelms any surrounding mass left in the stellar cavity; we verify this by considering a relatively low mass-loss rate of $10^{-6} M_{\odot}$ \citep{smi14}, combined with a wind velocity of 10 km/s, and a lower bound cavity size of 2 pc (to be calculated later). This results in a lower-bound for the mass in the surrounding medium:
\begin{equation} M_{wind} = \frac{R\dot{M}}{v} \approx 0.2 M_{\odot} \end{equation}
Similarly, an upper bound for high-mass stars can be approximated with a mass-loss rate of $10^{-4} M_{\odot}$/yr, a wind velocity of 300 km/s, and a cavity size of 20 pc, obtaining $M_{wind} \approx 6.5 M_{\odot}$. Therefore, we obtain a range of surrounding wind masses $0.2 \leq M_{wind} \leq 6.5$, compared to ejecta masses of $6.6 \leq M_{ej} \leq 118.6$ such that, $M_{ej} > M_{wind}$.

With very little
  medium in the way, the blast wave quickly traverses the cavity in a
  free expansion phase, and it does not slow down appreciably until it reaches the
  swept up shell.  Once the blast wave reaches the shell, there's
  enough material in the shell and the circumstellar medium to quickly
  slow down the evolution from a free expansion to a Sedov-like
  expansion.  Because the blast wave would spend very little time in
  the free expansion phase inside the wind-swept cavity (see eq.~\ref{eq:FEtime}), one would
  detect very few SNRs with sizes less than the wind-swept cavity
  size.  Therefore, the size of the wind-swept cavity presents a
  minimum size for the SNRs.

A consequence for the size distributions of SNRs in Sedov expansion
immediately follows: a minimum cut-off exists for SNR distributions
whose size is determined by the strength of the pre-explosion stellar
winds. An estimate of this cut-off can be found through a simple
order-of-magnitude calculation. Through ram pressure, the
  wind blows a bubble in the ISM.  Since the ram pressure declines
  with radius, but the ISM thermal pressure does not, at some point
  the wind ram pressure drops until it is roughly equal with the ISM
  thermal pressure.  This rough balance in pressures essentially sets
  the size of the wind-swept cavity. Recall that, for a fluid, the ram pressure
is $P \sim {\rho}v^2$ where in this case, $P$ is the pressure of the ISM and $v$ the velocity of the stellar wind. With the stellar mass-loss rate $\dot{M} = 4{\pi}{\rho}vR^2$, we can then substitute for the density and derive an estimate for the radius of the stellar cavity:
\begin{equation} R \sim \sqrt{\frac{\dot{M}v}{4{\pi}{\rho}}} \end{equation}

Thus, for mass-loss rates of $10^{-6} {M_\odot}$/yr to $10^{-4}
{M_\odot}$/yr \citep{smi14}, an ISM pressure of $P = 3000k_B$
K/cm$^3$, and a relatively-slow wind velocity of $v = 30$ km/s, one
obtains a range of cavity sizes from $\approx$ 2 to 20 pc. This is remarkably similar to the lower end of the SNR size
  distribution in Figure~\ref{fig:m31dist}.  Yet, we have already shown that the
  distribution of densities is also consistent with the WIM.  Either
  SNRs are biased to a narrow range of radii (because they trace the
  WIM) or they all have windblown cavities of order 2 to 20 pc.
  
Additionally, one might hope to use SN rates for M31 or M33 to assess
whether SNR catalogs are biased.  However, we find that the
current estimates for SN rates are far too uncertain to provide such a
constraint.  The most recent efforts characterize SN rates as a function of
 specific star formation rate (sSFR) \citep{gra15, bot16}.  With this correlation we can estimate SN rate for
 M31 and with a SNR lifetime we can estimate the number of SNRs.

The sSFR for M31 from recent estimates is $6.8\times10^{-12} \, {\rm M}_{\odot}^{-1}{\rm yr}^{-1}$ plus or
 minus about 20\%, which we calculate using M31's recent star formation history \citep{lew15} and stellar mass \citep{sic15} from the M31 Panchromatic
 Survey.  The correlation between type-II SN rate and sSFR is
 $\log(R_{\rm II}) = \log(A_s) B_s \log({\rm sSFR})$,
where $A_s = 7.9^{4}_{-3.4}$ and  $B_s = 1.33^{+0.41}_{-0.35}$ \citep{gra15}.  The
fraction of CCSNe that are II is $0.693 \pm 0.067$ \citep{li11,shi16}.

Given these estimates, the most likely SN rate for M31 is
0.0016 SNe/yr with an uncertainty in the range between 0.0005 and
0.0058 SNe/yr.  To calculate the number of expected SNRs, we need to know
the typical lifetime of a SNR.  A good guess would put this somewhere
between $10^4$ and $10^5$ years.  The geometric mean for this best
guess is $3 \times 10^4$ years.  Therefore, the best prediction for
the number of SNRs in M31 is 50.  However, our best estimates for the
uncertainty allows a range between 10 and 250 SNRs.  Clearly,
predictions for the number of expected SNRs is far too uncertain to
constrain whether current SNR catalogs are biased and missing SNRs.
In fact, historical trends suggest that every new attempt to find SNRs
yields yet more SNRs \citep{plu08,lon10,lee14}.

\section{Concluding Remarks}

We use the size distributions of SNRs to infer the properties of the
ISM. If we consider the simplest and most straightforward model, the
Sedov phase, then the narrow size distribution implies a narrow range
of densities. Either SNR surveys are biased towards the warm, ionized
medium, or the most obvious and simple model for SNR evolution is incorrect.

Though our statistical analysis is straightforward and simple, this
conclusion seems robust and may provide insight into SNR evolution and
observational biases.  Quite simply,
the distribution of SNR sizes is log normal and narrow.  The
log-normal character is consistent with the most common assumption
made about SNR evolution; the Sedov phase dominates the
visible phase of SNR evolution.  Even though there are only three
random variables in the Sedov radius, 
simple Monte Carlo simulations demonstrate that one can reproduce these log-normal
distributions for appropriately-sized SNR samples, less than $\sim$600
SNRs.  The puzzling result of these studies is that,
assuming a simple Sedov mode for the SNR blast wave, both data and
simulations reveal an unusually-small variance for the surrounding ISM
density.

Our primary motivation in inferring the density distribution is to
assess whether there is a bias in SNR catalogs.  Given our result, we are now in a position to address whether there is a bias in SNR catalogs.
 First of all, it is not clear whether the density distribution is
 actually narrow or if there are other mitigating factors; for example, SNR
 evolution may not be dominated by the Sedov phase. However, the
 independent results of two groups support our conclusion that the
 density distribution is narrow. \citet{gat16} simulated SNe in
 galactic simulations and found that most SNe explode in the same
 range of densities that we infer from observations. In addition,
 \citet{sar16} used radio emission to infer the radio
 emission lifetime and ambient densities (omitting inference from SNR sizes, in contrast to our method).  Both authors, in some
 fashion, estimated the number of SNRs below detection limits and ambient densities, with \citet{gat16} obtaining results similar to
 our own and \citet{sar16} obtaining mean densities nearly two orders of magnitude
 higher. We find that these results are a direct and simple
 consequence of the narrow diameter distribution in the context of
 Sedov evolution and, taken together, suggest that SNe tend to explode in the WIM. In this interpretation, either {\emph{all}} SNe explode in the WIM, or we are only seeing the ones that do.

\citet{sar16} analyzed a
 single M33 radio survey of SNRs, and they found that 30-40\%
 of actual SNRs fall below the detection limits of current radio
 surveys.  Taken at face value, their results suggest that current
 catalogs contain most of the SNRs in a galaxy.  This implies that most SNe explode in the WIM, and we are not missing the dominant fraction of SNRs.
 
Even if current catalogs contain most of the SNRs, the missing SNRs
could represent a bias against finding certain SNRs. For example, if the most massive stars exploded in HII regions, then these
might escape detection. Due to the scarcity of very massive stars, this
reduction could represent a small fraction in the total number of
SNRs. For example, if all progenitors above 35$M_{\odot}$ exploded in
HII regions, then this would result in only an approximate 14\% reduction in the total number of SNRs from our
chosen catalogs. Given our rough estimates, it is unlikely that we can
determine whether $\sim$10\% of SNRs are missing.  Therefore, there is a possibility that SNRs associated with the most massive progenitors are missing from catalogs.

In an attempt to constrain a mass dependence
scenario, we applied our inference technique as a function of
progenitor mass. \citet{jen14} published progenitor masses for the
SNRs that we consider; in figures ~\ref{fig:lowmass} and ~\ref{fig:highmass}, we infer the density distributions for the
high mass and low mass progenitors. We do not find a statistical
difference between low mass and high mass progenitors, thus suggesting
that if there is a bias it is largely independent of mass, or the mass
dependence bias does not dominate the SNR statistics.

Another possibility is that the evolution of SNRs in the Sedov phase
does not entirely determine the shape of the SNR size distribution.
Cavities blown by the progenitor's stellar winds may partially account
for the small variance in that they may provide
a minimum size for SNRs.

In summary, we find that a simple statistical inference of the SNR
size distribution provides an interesting constraint on the evolution
of SNRs.  Either SNe mostly explode in the WIM, there is a bias to
only {\emph{observing} SNRs that explode in the WIM, or the most obvious
assumption about SNR evolution---that their visibility is dominated by
the Sedov phase---is incorrect.  Although our simple analysis does
not definitively distinguish between these options, it does highlight
potential problems with using SNR catalogs to
understand SN physics.  None the less, SNRs remain important markers
of recent SNe, and so SNR catalogs continue to be some of the most
important catalogs for understanding SN physics.

To harness the full potential of SNR catalogs, we suggest the
following avenues of investigation.  First of all, we simply require
more data; identifying more SNRs will help to further constrain the distribution of SNR
sizes. If we can obtain thousands of SNRs, then one will be able to
model the distribution beyond log-normal, allowing us to infer more
parameters than the mean and variance.  This will in turn provide more
constraints on theoretical models for SNR evolution.  Furthermore,
these catalogs should be defined using the most robust SNR
identification techniques.  Second, we require more
data about the environments of SNRs that are independent of SNR
evolutionary models.  Finally, we need better SNR evolutionary models and the best way to
achieve this would be to better understand the evolution of individual SNRs.  This will
require a concerted effort between detailed observations of local SNRs
and detailed evolutionary modeling.  Together, more data and better
theoretical modeling will enable us to constrain potential biases in
SNR catalogs, allowing us to better use them to infer the physics of
SNe.
\section*{Acknowledgments}

We thank Carles Badenes for suggesting that we investigate the size distributions of supernova remnants.  This material is based upon work supported by the National Science Foundation under Grant No. 1313036.










\label{lastpage}
\end{document}